\documentclass[%
superscriptaddress,
 reprint,
 amsmath,amssymb,
 aps,
]{revtex4-2}

\usepackage{graphicx}
\usepackage{dcolumn}
\usepackage{bm}
\usepackage{hyperref}

\begin{document}

\preprint{APS/123-QED}

\title{Short-time critical dynamics in the classical cubic dimer model}

\newcommand{\SYSU}{School of Physics, Sun Yat-sen University, Guangzhou 510275, China}
\newcommand{\GUANGDONGLAB}{Guangdong Provincial Key Laboratory of Magnetoelectric Physics and Devices, \SYSU}
\newcommand{\WESTLAKEPHY}{Department of Physics, School of Science and Research Center for Industries of the Future, Westlake University, Hangzhou 310030, China}
\newcommand{\WESTLAKEINST}{Institute of Natural Sciences, Westlake Institute for Advanced Study, Hangzhou 310024, China}

\author{Hu-Xiao Peng}
\affiliation{\SYSU}
\author{Zheng Yan}
\email{zhengyan@westlake.edu.cn}
\affiliation{\WESTLAKEPHY}
\affiliation{\WESTLAKEINST}

\author{Shuai Yin}
\email{yinsh6@mail.sysu.edu.cn}
\affiliation{\SYSU}
\affiliation{\GUANGDONGLAB}



\date{\today}

\begin{abstract}
The classical dimer model on the cubic lattice hosts a columnar ordered phase and a disordered Coulomb phase, separated by a continuous phase transition that lies beyond the conventional Landau-Ginzburg-Wilson paradigm. While its equilibrium critical properties have been extensively studied, the nonequilibrium critical dynamics of this model—particularly in the short-time regime—remains largely unexplored. In this work, we investigate the short-time critical dynamics near the transition using large-scale Monte Carlo simulations. By quenching the system from both ordered and disordered initial states with vanishing initial correlation length, we analyze the scaling behaviors of the order parameter and its time correlation function in the short-time stage. From these scaling behaviors, we accurately determine the critical temperature $T_c = 0.672(1)$ and the static critical exponent $\beta/\nu = 0.581(5)$ according to the scaling theory of the short-time dynamics. These results are in excellent agreement with previous equilibrium studies. Moreover, we extract the dynamic critical exponent $z = 1.92(1)$ and, notably, find a \emph{negative} critical initial slip exponent $\theta = -1.052(5)$. This unusual negative value contrasts sharply with the positive $\theta$ typically observed in conventional critical dynamics. We attribute this anomalous behavior to the combined effects of the emergent SO(5) symmetry at criticality and the local U(1) gauge constraint (Gauss law), which enforces a conserved diffusive dynamics and enhances fluctuations in the short-time regime. Our results provide the first comprehensive characterization of nonequilibrium short-time criticality in the three-dimensional dimer model, shedding new light on the universal dynamical features of phase transitions beyond the Landau-Ginzburg-Wilson framework.

\noindent \textbf{Keywords:} classical dimer model, nonequilibrium critical dynamics, short-time critical dynamics, Monte Carlo simulation, critical exponents

\end{abstract}

\maketitle


\section{Introduction}\label{sec:introduction}

Understanding critical phenomena has been one of the central topics in condensed matter physics for decades. In recent years, intensive efforts have been devoted to investigations on continuous phase transitions beyond the Landau-Ginzburg-Wilson (LGW) paradigm. A typical example is the deconfined quantum criticality, which describes the putative continuous phase transition between two ordered phases with incompatible broken symmetry~\cite{doi:10.1126/science.1091806,PhysRevB.70.144407,PhysRevLett.98.227202,
PhysRevLett.100.017203,
Jiang_2008,
PhysRevB.80.180414,
PhysRevB.82.155139,
PhysRevLett.104.177201,
PhysRevLett.111.087203,
PhysRevB.88.220408,
PhysRevLett.110.185701,
PhysRevX.5.041048,
PhysRevLett.115.267203,
PhysRevB.92.184413,zhang2018continuous}. Furthermore, although the deconfined quantum criticality was originally proposed in quantum phase transitions, it was shown that it also has its classical counterparts~\cite{PhysRevLett.101.050405,PhysRevLett.101.155702,PhysRevLett.101.167205,PhysRevB.80.134413,PhysRevB.80.045112,annurev-conmatphys-070909-104138}.

The classical dimer model is one of the typical models which hosts such unconventional phase transitions~  \cite{KASTELEYN19611209,kasteleyn1963dimer,PhysRev.132.1411,Henley1997,PhysRevLett.91.167004,PhysRevLett.97.030403}. In two dimensions, the dimer model on bipartite lattice undergoes the Kosterlitz-Thouless transition from a columnar phase at low temperatures to a disordered critical phase at high temperatures~\cite{PhysRevLett.94.235702,PhysRevE.74.041124,PhysRevB.76.134514,dabholkar2023classical,dabholkar2022reentrance}. In contrast, the three-dimensional ($3$D) bipartite lattice dimer model exhibits a different type of phase transition. Although the system also evolves from a disordered phase at high temperatures to a columnar ordered phase at low temperatures, the transition in $3$D is continuous and cannot be described by the framework of conventional LGW theory~\cite{PhysRevLett.97.030403,PhysRevB.78.100402,PhysRevB.82.014429}.

It was shown that the phase transition in $3$D dimer model can be characterized by a field theory that features two complex matter fields coupled to a non-compact U(1) gauge field~\cite{PhysRevLett.101.155702,PhysRevLett.101.167205,PhysRevB.80.134413,PhysRevB.80.045112,annurev-conmatphys-070909-104138}. In addition, it has been demonstrated that the transition in the three-dimensional cubic dimer model exhibits an emergent SO(5) symmetry at criticality~\cite{PhysRevLett.122.080601}. In addition, critical exponents obtained from numerical simulations do not match those of conventional LGW universality class~\cite{PhysRevLett.97.030403,PhysRevB.82.014429}. 

These intriguing properties have made this model attract extensive research attention. However, previous studies mainly focus on the equilibrium properties of this model~\cite{PhysRevB.82.014429,PhysRevLett.104.045701,PhysRevB.89.014404}. While nonequilibrium critical dynamics has been explored in certain quantum systems, which share the similar effective field theory~\cite{PhysRevLett.128.020601,PhysRevB.105.104420}, the nonequilibrium properties for the classical dimer models remain largely unknown. 

Among the diverse forms of nonequilibrium processes, relaxation dynamics stands out as a representative prototype~\cite{janssen1989new,PhysRevB.40.304,Z-B.Li_1994,SCHULKE199681,
LUO1998383,PhysRevLett.74.3396,SCHULKE1995295,PhysRevLett.77.679,OKANO1997727,PhysRevE.58.4242,ZHENG1999338,ZHENG200080,PhysRevLett.81.180}. It was shown that universal dynamic scaling behaviors not only appear in the long-time stage~\cite{RevModPhys.49.435,Folk_2006}, but also emerge in the short-time stage~\cite{janssen1989new,PhysRevB.40.304}, characterized by the critical initial slip exponent $\theta$, in addition to the usual dynamic exponent $z$. In usual classical systems, $\theta$ is positive~\cite{janssen1989new}. Furthermore, short-time dynamics offers a high-efficiency strategy for extracting critical points and critical exponents without equilibrating the system, thus conserving computational resources while circumventing the impacts of critical slowing down~\cite{janssen1989new,PhysRevB.40.304,Z-B.Li_1994,SCHULKE199681,
LUO1998383,PhysRevLett.74.3396,SCHULKE1995295,PhysRevLett.77.679,OKANO1997727,PhysRevE.58.4242,ZHENG1999338,ZHENG200080,PhysRevLett.81.180}. In addition, the short-time dynamics was generalized to the quantum phase transitions~\cite{PhysRevB.89.144115,PhysRevE.90.042104,PhysRevLett.123.170606,PhysRevB.102.104425,PhysRevLett.128.020601,PhysRevB.104.214108,yu2025nonequilibriumdynamicsdiracquantum,PhysRevB.109.134309,PhysRevB.109.184303,doi:10.1126/sciadv.adz4856,shen2025universalentanglementgrowthimaginary,zhang2025magneticordernovelquantum}. However, to the best of our knowledge, the dynamic critical exponent $z$ and the critical initial slip exponent $\theta$ have not been determined for the three-dimensional classical dimer model in previous studies. 

Given the significance of the dimer model and the universality of relaxation critical dynamics, investigations into the relaxation dynamics of phase transitions in the dimer model are imperatively required. In this paper, we explore the short-time dynamics of the $3$D classical cubic dimer model starting from both ordered and disordered initial states via using the Monte Carlo simulation. By focusing on short-time dynamics of the square of the order parameter component $N_x^2$ for different initial states, we retrieve both the critical temperature $T_c$ and the static critical exponent $\beta/\nu$, which agree well with equilibrium finite-size scaling results; additionally, we successfully derive the dynamic critical exponent $z=1.92(1)$. Moreover, we also determine the critical initial slip exponent $\theta$ from the time-correlation function $Q(t)$ of the order parameter starting from disordered initial states. Remarkably, we find that $\theta$ is negative here, in sharp contrast to the usual cases in which $\theta$ is positive. We attribute this to the emergent symmetry or the local constraint induced by the Gauss law at the critical point of this model.

The remainder of this paper is organized as follows. In Sec.~\ref{sec:model}, we introduce the classical cubic dimer model and the relevant observables, and describe the algorithm for the study. In Sec.~\ref{sec:scaling}, we briefly review the short-time dynamic scaling and show the expected scaling relations for ordered and disordered initial conditions. The numerical results for the short-time dynamics are presented and analyzed in Sec.~\ref{sec:results}, where the critical temperature and the critical exponents are determined through scaling analyses. Finally, Sec.~\ref{sec:summary} is devoted to a summary of our main results.

\section{Model and numerical method}\label{sec:model}

The dimer model studied here is defined on a cubic lattice of linear size $L$ (total number of sites $N = L^3$), where the lattice links are occupied by hard-core dimers. By defining $d_{\mu}(\vec{r}) \in [0,1]$ as the occupation number between the site $\vec{r}$ and $\vec{r} + \vec{\delta}_{\mu}$ ($\vec{\delta}_{\mu}$ is a unit vector in $\mu$-direction), the number of dimers at the site $\vec{r}$ is given by $n(\vec{r}) = \sum_{\mu} \left[d_{\mu}(\vec{r}) + d_{\mu}(\vec{r} - \vec{\delta}_{\mu})\right]$. Only dimer configurations $\mathcal{C}$ that obey the closed-packing condition $n(\vec{r}) = 1$, i.e., one dimer per site, contribute to the partition function. 

\begin{figure}[]
    \includegraphics[width=0.48\textwidth]{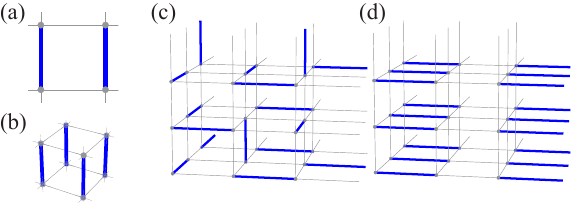}
    
    \caption{\label{fig:example1}%
        Two types of interactions and two types of initial states in the dimer model. (a) Two parallel dimers on the plane contribute energy \(v_{2}\). (b) Four parallel dimers on a unit cube contribute energy \(v_{4}\). In this study, we fixed \(v_{2} = -1\) and \(v_{4} = 10\). (c) The disordered initial state with $N_x = 0$, typical of a Coulomb phase. (d) The completely ordered initial state with $N_x = 1$, representing a columnar phase.}
\end{figure}

The partition function is defined as $\mathcal{Z}\equiv \sum_{\mathcal{C}}\mathrm{e}^{-\beta E_{\mathcal{C}}}$, wherein the energy of an allowed configuration $\mathcal{C}$ is given by~\cite{PhysRevB.82.014429}
\begin{eqnarray}
E_{\mathcal{C}}=v_{2}N_{2}+v_{4}N_{4},
\end{eqnarray}
where $N_2$ is the number of plaquettes of the lattice that contain two parallel dimers, and $N_4$ is the number of unit cubes sustaining four parallel dimers, both illustrated in Fig.~\hyperref[fig:example1]{1}(a)-(b). 

In this study, we restrict ourselves to the case of attractive plaquette interactions $v_{2}$ while the cubic interaction $v_{4}$ is repulsive. We focus on the strongly frustrated condition to avoid any crossover effects~\cite{PhysRevB.82.014429}. In the remainder of this study, we set $v_{2} = -1$ and $v_{4} = 10$ which are the same as a set of parameters in Ref.~\cite{PhysRevB.82.014429}. Therefore, the results obtained in the following contents will also be compared with those in Ref.~\cite{PhysRevB.82.014429}.

The system undergoes a continuous transition from the disordered Coulomb phase to the ordered phase~\cite{PhysRevB.82.014429}. At the critical point, an emergent SO(5) symmetry appears and the order parameter has five equivalent components~\cite{PhysRevLett.122.080601}. In this work, we focus on a single component $N_x$ of the order parameter. Here $N_x$ denotes the thermal-averaged $x$-component of the order parameter, defined as
\begin{equation}
N_x
\equiv \frac{2}{L^{d}}
\left\langle
 \sum_{\vec{r}} (-1)^{r_x} d_x(\vec{r})
\right\rangle ,
\end{equation}
where $r_x$ is the $x$-component of $\vec{r}$ and $\langle \cdots \rangle$ represents the thermal average. In the thermodynamic limit, $N_x$ is zero in the disordered Coulomb phase and nonzero in the columnar ordered phase.

Since the order parameter $N_x$ averages to zero in finite-size systems, critical properties cannot be directly extracted from $N_x$. It is therefore convenient to consider the square of the order parameter, defined as
\begin{equation}
N_x^{2}
\equiv \frac{4}{L^{2d}} 
\left\langle
\left(
 \sum_{\vec{r}} (-1)^{r_x} d_x(\vec{r})
\right)^2
\right\rangle .
\label{Nxsquare}
\end{equation}
By analyzing the short-time scaling behavior of $N_x^2$, one can obtain both static critical exponent $\beta/\nu$ and dynamic critical exponent $z$~\cite{PhysRevLett.74.3396,SCHULKE1995295}.

Moreover, the short-time critical dynamics shows that the time-correlation function contains additional information about the universal critical initial slip in the early-time stage~\cite{janssen1989new}. The critical initial slip behavior is characterized by the critical initial exponent $\theta$, which can be readily obtained via the total time-correlation of the order parameter, $Q(t)$, proposed in Ref.~\cite{PhysRevE.58.4242}. Specifically, here $Q(t)$ reads
\begin{equation}
Q(t)
\equiv \frac{2}{L^{d}}
\left\langle
\left(
\sum_{\vec{r}} \sum_{\vec{r}'}(-1)^{r_x+r'_x}d_x(\vec{r},t) d_x(\vec{r}',0)
\right)
\right\rangle .
\end{equation}


We employ the standard Metropolis algorithm~\cite{BinderHeermann2019,LandauBinder2021} with local updates to simulate the nonequilibrium dynamics. It has
been established that the Metropolis Monte Carlo dynamics falls within the Model A universality class, and is straightforward to implement in experiments~\cite{RevModPhys.49.435,Folk_2006}. Specifically, we randomly select a plaquette containing two parallel dimers, as illustrated in Fig.~\hyperref[fig:example1]{1}(a), and update the configuration by flipping these dimers to the two adjacent links that are initially unoccupied. The time unit is defined as a Monte Carlo sweep through the lattice.

We study the nonequilibrium short-time dynamics from both the disordered initial state and the completely ordered initial state. Both initial states are illustrated in Fig.~\hyperref[fig:example1]{1}(c)-(d). The ordered state is prepared straightforwardly by placing the dimers on alternating links along the $x$ axis, as shown in Fig.~\hyperref[fig:example1]{1}(d). The disordered initial states are obtained by evolving this ordered configuration at a sufficiently high temperature until the order parameter vanishes and becomes stationary. In this way, both initial states belong to the same topological sector~\cite{PhysRevLett.91.167004}.

\section{Scaling for the short-time dynamics}\label{sec:scaling}

In addition to the universal behavior in the long-time regime~\cite{RevModPhys.49.435,10.1093/acprof:oso/9780198509233.001.0001},
systems near criticality can exhibit universal relaxation dynamics in the macroscopic short-time stage, which sets in immediately after a microscopic timescale $t_{\mathrm{mic}}$~\cite{janssen1989new}. It was shown that starting from an uncorrelated state with initial order parameter $m_0$, the scale transformation of a macroscopic observable $O$ is given by~\cite{janssen1989new,PhysRevLett.77.679},
\begin{equation}
O(t,\tau,L,N_{x0})
= b^{-x}
O(b^{-z}t,\; b^{1/\nu}\tau,\; b^{-1}L,\; b^{x_0} N_{x0}),
\label{observable}
\end{equation}
in which $b$ is the rescaling factor, $t$ denotes the relaxation time, $\tau$ is the reduced temperature, $L$ is the linear system size, $x$ is the scaling dimension of $O$, $\nu$ is the correlation length exponent, $z$ is the dynamic critical exponent, and $x_0$ is the dimension of the initial order parameter $N_{x0}$ and satisfies $x_0 = \theta z + \beta/\nu$. For instance, the square of the order parameter $N_x^2$ changes as~\cite{janssen1989new,PhysRevLett.77.679},
\begin{equation}
N_x^2(t,\tau,L,m_0)
= b^{-2\beta/\nu}
N_x^2(b^{-z}t,\; b^{1/\nu}\tau,\; b^{-1}L,\; b^{x_0} N_{x0}),
\label{Nx2_scaling}
\end{equation}
under scale transformation with the rescaling factor $b$.

For the uncorrelated disordered initial state, $N_{x0}=0$, which corresponds to the disordered fixed point of $N_{x0}$, the dynamic scaling form of $N_x^2$ at the critical point follows by setting the rescaling factor $b=L$ in Eq.~(\ref{Nx2_scaling})~\cite{janssen1989new,PhysRevLett.77.679},
\begin{equation}
N_x^2(t,L)
= L^{-2\beta/\nu}f_{N_x}\left(tL^{-z}\right),
\label{disorder_FSS}
\end{equation}
in which $f_{N_x}$ is the scaling function.

In the short-time regime, the correlation length is smaller than the system size. Accordingly, the definition of $N_{x}^2$ in Eq.~\eqref{Nxsquare} indicates that $N_x^2 \propto L^{-d}$. In this situation, the scaling function $f_{N_x}$ should satisfy~\cite{Albano_2011},
\begin{equation}
f_{N_x}(t L^{-z}) \propto (t L^{-z})^{\kappa}.
\label{kappa}
\end{equation}
Combining Eq.~(\ref{kappa}) and Eq.~\eqref{disorder_FSS}, the fact that $N_x^2 \propto L^{-d}$ requires that,
\begin{equation}
\kappa = (d-2\beta/\nu)/z,
\end{equation}
leading to the short-time power-law scaling behavior~\cite{PhysRevB.40.304,SCHULKE1995295,Albano_2011},
\begin{equation}
N_x^2(t,L) \propto L^{-d}\, t^{(d-2\beta/\nu)/z}.
\label{disorder}
\end{equation}

In addition, for the disordered initial state with $N_{x0}=0$, at the critical point $\tau=0$, the time-correlation function $Q$ obeys the dynamic finite-size scaling form~\cite{PhysRevE.58.4242},
\begin{equation}
Q(t,L)
= L^{\theta z}\,
f_{Q}\!\left(tL^{-z}\right),
\label{Q_FSS}
\end{equation}
in which $f_{Q}$ is also the scaling function. In the short-time stage $t \ll L^{z}$, the correlation length is far smaller than the system size; accordingly, $Q(t)$ is expected to be size-independent. In this way, Eq.~(\ref{Q_FSS}) reduces to a power-law scaling relation~\cite{PhysRevE.58.4242},
\begin{equation}
Q(t) \propto t^{\theta}.
\label{correlation}
\end{equation}


For the completely ordered initial state, in which $N_{x0}=1$, corresponding to the ordered fixed point of $N_{x0}$, in the short-time regime and for sufficiently large system sizes, $N_x^2$ exhibits a power-law decay~\cite{janssen1989new,PhysRevLett.77.679,ZHENG1999338},
\begin{equation}
N_x^{2}(t) \propto t^{-2\beta/\nu z},
\label{order}
\end{equation}
which is independent of $L$ since the correlation length is much smaller than $L$.

For notational convenience, in the following discussion we denote the order parameter $N_x^2$ by $N_d^2$ for disordered initial conditions and by $N_o^2$ for ordered initial conditions.

\section{Short-time dynamics in the classical dimer model}\label{sec:results}


First, we locate the critical temperature $T_c$ via the short-time dynamics. From the scaling relation~\eqref{disorder}, at $T_c$, $N_d^{2}(t)$ should evolve obeying a power function of the time~\cite{SCHULKE199681,PhysRevLett.81.180}; while away from the critical point, $N_d^{2}(t)$ will deviate from the power function. This demonstrates the feasibility of determining the critical point from the short-time dynamics of $N_d^{2}(t)$.

\begin{figure}[]
    \centering
    \includegraphics[width=0.96\linewidth]{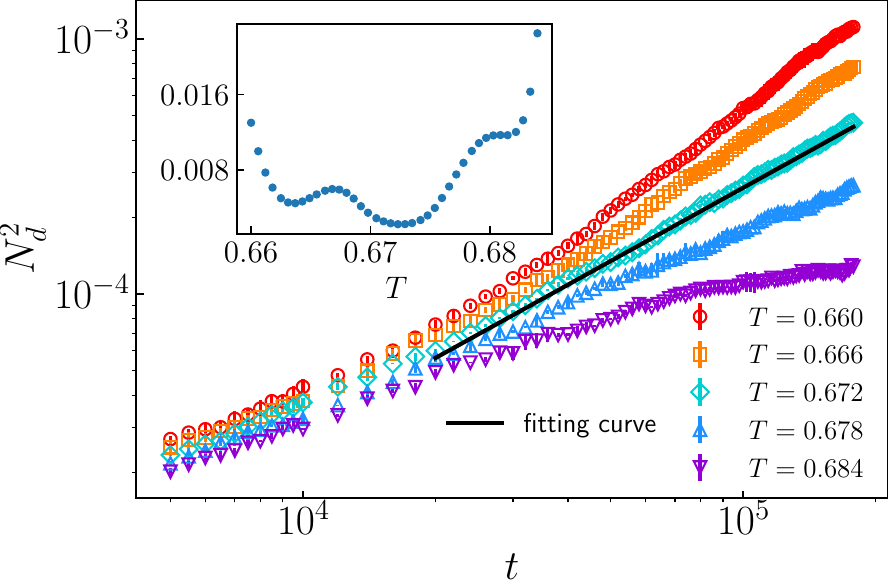}\\
    
    \caption{
        Estimation of the critical point of the $3$D dimer model. The time evolution of $N_d^{2}(t)$ starting from the disordered state for different temperatures. The solid line shows the power-law fit of $N_d^{2}(t)$ versus $t$ in the time range $t\in [2\times 10^4,2\times 10^5]$ at $T_c=0.672$. The inset displays the squared deviation of $N_d^{2}(t)$ from the power law behavior. The system size is chosen as $L = 100$. Log-log coordinates are used for the main figure.
   }%
   \label{fig:point}%
   \phantomsection
\end{figure}

In practice, the critical point is determined by searching for the minimum squared deviation of the slope of the curves of $N_{x}^2$ versus $t$ in log-log coordinates for different temperatures. Here, the system size is chosen as $L=100$, for which we have confirmed the effect of finite size on the position of the critical point can be ignored. The simulations are carried out with five temperatures $T=0.660$, $0.666$, $0.672$, $0.678$ and $0.684$ as shown in Fig.~\hyperref[fig:point]{\ref{fig:point}}. Data within the microscopic time scale \( t_{\text{mic}} \), which are dependent on microscopic details, are not included. The values of $N_d^{2}(t)$ at temperatures between $T=0.660$ and $0.684$ are obtained by quadratic interpolation. Based on these interpolated data, the slopes of $N_d^2$ as a function of time $t$ are fitted for different temperatures. Specifically, for each temperature, data for $N_d^2$ in the time interval between $2\times 10^4$ and $2\times 10^5$ are partitioned into four subintervals, from which local slopes are extracted. The squared deviation of these slopes in subintervals from their averaged value is then calculated to quantify the deviation from a power-law behavior. In the inset of Fig.~\hyperref[fig:point]{\ref{fig:point}}, the squared deviation is shown for different temperatures. The minimum position in this plot yields that the critical point is near $T=0.672$. To estimate the statistical error of the critical temperature $T_c$, we first fit the curve of $N_d^2$ versus $t$ in the range $t\in [2\times 10^4,2\times 10^5]$ at $T = 0.672$, i.e., the solid line in the main plot of Fig.~\hyperref[fig:point]{\ref{fig:point}}. Then we perform fitting on the interpolated data for different temperatures around $T=0.672$ over the same time range and the minimum deviation of the temperature at which the fitted slope falls outside the error range of the slope for $T = 0.672$ provides an estimate of the error. Accordingly, we obtain $T_c = 0.672(1)$. This value is consistent with those reported in previous studies, $T_c = 0.6718(2)$ ~\cite{PhysRevLett.122.080601} and $0.672(1)$~\cite{PhysRevB.82.014429}, wherein the usual equilibrium finite-size scaling methods were employed.

\begin{figure}[b]
    \centering
    \includegraphics[width=0.7\linewidth]{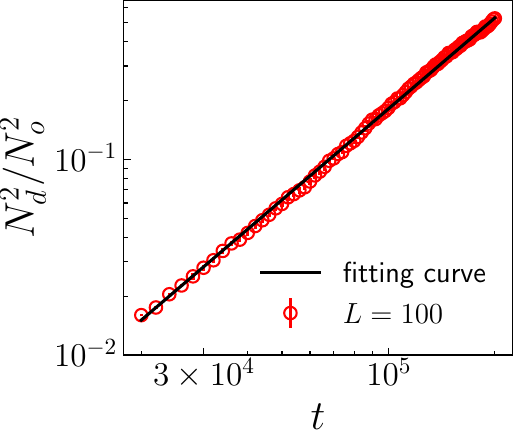}
    \caption{Short-time dynamics of the ratio \( N_d^2/N_o^2 \) at the critical point $T_c = 0.672$ for system size $L=100$. The dotted curve represents the Monte Carlo results, while the solid line corresponds to the fitted curve with a power function. Log-log coordinates are used.}
    \label{fig:dis_order}
\end{figure}

Then we determine the dynamic exponent $z$. We consider the evolution of the ratio $N_d^{2}/N_o^{2}$~\cite{DASILVA2002325,Fernandes_2006}. According to the scaling relations~\eqref{disorder} and \eqref{order} for $N_d^{2}$ and $N_o^{2}$, respectively, in the short-time scaling range, $N_d^{2}/N_o^{2}$ should satisfy $N_d^{2}/N_o^{2} \propto t^{d/z}$~\cite{DASILVA2002325,Fernandes_2006}. Fig.~\hyperref[fig:dis_order]{\ref{fig:dis_order}} shows the numerical results of $N_d^{2}/N_o^{2}$ in the time range $t \in [2 \times 10^{4},\, 2 \times 10^{5}]$ at $T_c=0.672$. The power fit gives $d/z = 1.56(1)$. Substituting $d=3$, we obtain the dynamic critical exponent as $z = 1.92(1)$. For comparison, in the three-dimensional Ising model with Model A dynamics the dynamic exponent is typically reported as $z = 2.04(3)$~\cite{PhysRevB.43.6006}. The smaller value obtained here may be related to the local constraint in the dimer model.
\begin{figure*}
        

    \includegraphics[width=\textwidth]{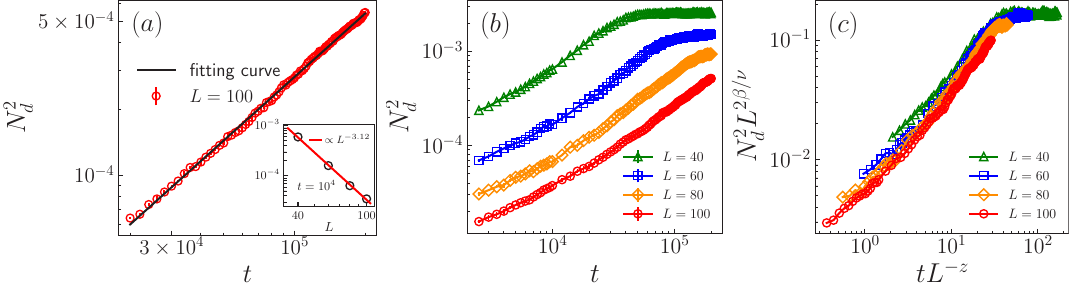}

    \caption{Short-time dynamics of the classical dimer model quenched from a disordered state to the critical point for various system sizes. (a) $ N_d^{2}$ plotted versus time in log-log scale for system size $L=100$ . The symbols represent the Monte Carlo results, while the solid line corresponds to the fitted result. The inset shows a log-log plot of $ N_d^{2}$ as a function of system size $L=40,60,80,100$ at a fixed time $t=10^4$, where the solid line represents a power-law fit yielding $N_d^{2} \propto L^{-3.12}$, consistent with the expected $ L^{-d} $ scaling with $ d = 3 $.  (b) Time dependence of $ N_d^2 $ for $L=40$ to $100$. (c) After rescaling $N_d^2$ and $t$ as $N_d^2L^{2\beta/\nu z}$ and $tL^{-z}$, all rescaled curves collapse well. Log-log coordinates are used.}
    \label{fig:result1}
\end{figure*}

\begin{figure*}
        


     \includegraphics[width=\textwidth]{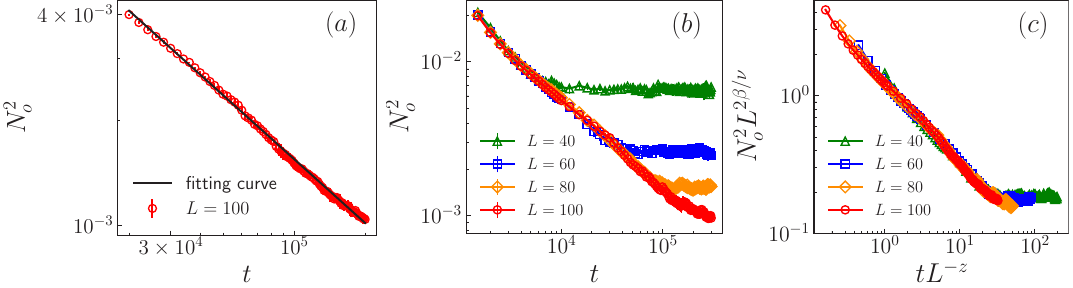}
    \caption{Short-time dynamics of the classical dimer model quenched from an ordered state to the critical point for 
    various system sizes. (a) \( N_o^{2}\) plotted versus time in log-log scale for system size $L=100$ . (b) Time dependence of \( N_o^2 \) for $L=40$ to $100$. (c) After rescaling $N_o^2$ and $t$ as $N_o^2L^{2\beta/\nu z}$ and $tL^{-z}$, all curves collapse well. Log-log coordinates are used.}
    \label{fig:result2}
\end{figure*}

With the dynamic critical exponent $z$ determined, the static critical exponent $\beta/\nu$ can be extracted from the scaling behaviors of $N_d^{2}$ and $N_o^{2}$ using the scaling relations~\eqref{disorder} and \eqref{order}~\cite{PhysRevLett.74.3396}. For the disordered initial state, the inset of Fig.~\hyperref[fig:result1]{4}(a) shows $N_d^{2}$ as a function of the system size $L$ at a fixed time $t = 10^{4}$. A power-law fit yields $N_d^{2} \propto L^{-3.12}$ with the exponent close to $d=3$, thereby verifying $N_d^{2} \propto L^{-d}$, as shown in Eq.~\ref{disorder}. Then, in the main plot of Fig.~\hyperref[fig:result1]{4}(a), by fitting the slope of the curve of $N_d^2$ versus $t$ for $L=100$ in log-log coordinates, we find that $(d-2\beta/\nu)/z = 0.957(2)$, leading to $\beta/\nu = 0.581(5)$ with $z = 1.92(1)$ and $d = 3$. Fig.~\hyperref[fig:result1]{4}(b) shows the short-time dynamics of $N_d^2$ for different system sizes. In particular, in the time range $t\in [2\times 10^4,2\times 10^5]$, the curves log-log coordinates for $L=100$ and $L=80$ are almost parallel, demonstrating that the exponent estimated above is almost independent of size. Additionally, by rescaling $N_d^2$ and $t$ as $N_d^2L^{2\beta/\nu z}$ and $tL^{-z}$, the rescaled curves collapse well, as shown in Fig.~\hyperref[fig:result1]{4}(c). This collapse indicates that the short-time dynamics of $N_d^{2}$ obeys the scaling form given in Eq.~\eqref{disorder_FSS}, and supports the estimated critical exponents $\beta/\nu$ and $z$.

Similarly, for the ordered initial state, the slope of the curve of $N_o^2$ versus $t$ in log-log coordinates for $L=100$ yields $-2\beta/\nu z = -0.601(3)$, leading to $\beta/\nu = 0.577(4)$, as shown in Fig.~\hyperref[fig:result2]{5}(a). This value is close to the one obtained from the disordered case. In addition, Fig.~\hyperref[fig:result2]{5}(b) shows that the evolution of $N_o^2$ is almost independent of the system size in the short-time stage. By performing the rescaling for $ N_o^2 $ and $ t $ as $ N_o^2 L^{2\beta/\nu z} $ and $ t L^{-z} $, respectively, the curves collapse well, as illustrated in Fig.~\hyperref[fig:result2]{5}(c), which not only supports the estimated exponent $\beta/\nu$ and $z$ but also verifies the dynamic scaling form Eq.~\eqref{disorder_FSS}. 

\begin{figure}
\includegraphics[width=0.48\textwidth]{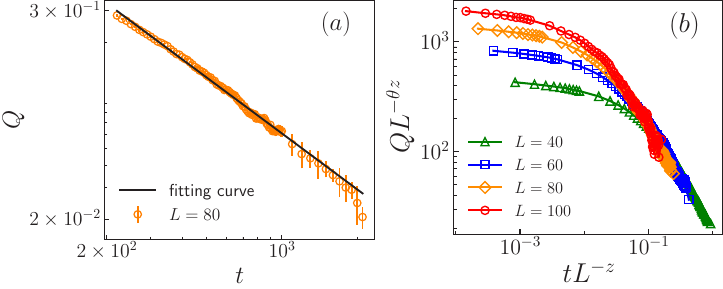}
\caption{The time correlation of the total
order parameter $Q$ starting from a disordered state at the critical point.
(a) Time dependence of $Q$ for system size $L=80$. Since the statistical fluctuations for $L=100$ are too large to allow for a reliable fit, we therefore use $L=80$ in our analysis.
(b) After rescaling $Q$ and $t$ as $QL^{-\theta z}$ and $tL^{-z}$, all curves collapse well. Log-log coordinates are used.}
\label{fig:result3}
\end{figure}

Moreover the values of $\beta/\nu$ obtained via the short-time dynamics from both the ordered and disordered initial states are consistent with previous results. In particular, combining the results $\eta = 0.25(3)$ and $\nu = 0.63(4)$ presented in Ref.~\cite{PhysRevB.82.014429} and the scaling laws, one obtains $\beta = 0.40(7)$, leading to $\beta/\nu = 0.6(1)$, which is consistent with our present results within the error bar.

Besides these usual long-time critical exponents, the short-time critical dynamics has an additional critical initial slip exponent $\theta$, which can be extracted from the scaling behavior of the time-correlation function, as shown in Eq.~(\ref{correlation}). Fig.~\hyperref[fig:result3]{6}(a) shows the evolution of $Q$ for $L=80$ in the time range $t\in[2.2\times 10^2,2.2\times 10^3]$. Power fit of the data yields the exponent $\theta = -1.052(5)$. In addition, by rescaling $Q$ and $t$ as $Q L^{-\theta z}$ and $t L^{-z}$ with the exponents $\theta$ and $z$ obtained, the rescaled curves collapse well after a transient time, as shown in Fig.~\hyperref[fig:result3]{6}(b), which not only confirms the value of $\theta$ but also verifies the scaling form Eq.~(\ref{Q_FSS}).

        



A striking observation is that $\theta$ takes a negative value here, which stands in contrast to the usual cases in the LGW transitions. When initial states are random and possess a small residual initial order parameter $m_0$, the order parameter’s evolution is regulated by two competing effects: critical fluctuations induce its decay, while domain formation arises because the residual $m_0$ tends to orient the neighboring order parameter toward the ordered state. In usual cases, the positive $\theta$ indicates that the latter element dominates and the order parameter, say $m$, will increase as $m\sim m_0 t^\theta$~\cite{janssen1989new}. The reason is that in the initial stage, there are no long-range correlations, so the system can be described by mean-field theory. In a typical LGW phase transition, critical fluctuations cause the actual critical temperature to be lower than its mean-field counterpart. Thus, in the short-time stage, the system behaves as if it were in an ordered phase, making domain formation the dominant factor in early evolution. 

In contrast, in the present case, the critical point has emergent SO($5$) symmetry, which describes the rotation symmetry between five components of critical modes in both sides of the critical point. These modes include not only the three components of the order parameter—whose fluctuations tend to drive the critical point below its mean-field value—but also two additional components that emerge in the disordered phase and drive the critical point in the opposite direction. The emergent SO($5$) demonstrates that the strength of fluctuations on both sides of the critical point is essentially the same. Therefore, critical fluctuations cannot cause a qualitative change in the real critical point compared to the mean-field critical point. Accordingly, the domain formation cannot govern the short-time dynamics, resulting in a negative $\theta$. 

The behavior that the critical point does not shift to values smaller than its mean-field value due to such additional degrees of freedom is analogous to that observed at Dirac criticality, wherein the order parameter fluctuations drive the critical point below its mean-field value, but the fermion fluctuations play opposite roles. It was shown that when the fermion fluctuations dominate, a negative $\theta$ also appears~\cite{PhysRevLett.123.170606,yu2025nonequilibriumdynamicsdiracquantum}.

Another possible reason for the reduction of $\theta$ is the presence of the Gauss law in the dimer model~\cite{PhysRevLett.101.155702,PhysRevLett.101.167205,PhysRevB.80.134413,PhysRevB.80.045112,annurev-conmatphys-070909-104138}. This constraint is equivalent to a local conservation law for the gauge charge, which can fundamentally alter the dynamics of the system. Such a constraint can hinder the local rearrangements required for the growth of ordered domains in the early-time regime. Consequently, the amplification of a small initial order parameter is suppressed. As reported in previous literature~\cite{Taeuber2014}, which shows that a globally conserved order parameter suppresses critical initial slip, we thus speculate that local constraints may do the same, resulting in a negative $\theta$. 

It should be noted that these two factors are independent, and we cannot rule out either. We expect to further verify our conclusions in similar models without emergent symmetries or local constraints in future work.

\section{Summary}\label{sec:summary}
In conclusion, we have systematically investigated the short-time critical dynamics of the three-dimensional classical cubic dimer model using Monte Carlo simulations. By employing the short-time scaling method, we have obtained a comprehensive set of critical parameters without requiring full equilibration of the system, thereby offering an efficient alternative to conventional equilibrium studies.

We have accurately determined the critical temperature $T_c = 0.672(1)$, which is in excellent agreement with values previously reported in the literature. The static critical exponent ratio $\beta/\nu = 0.581(5)$, extracted from the short-time power-law scaling of the order parameter, is also consistent with equilibrium results, validating the reliability of the dynamical approach. Furthermore, we have computed the dynamic critical exponent $z = 1.92(1)$, which governs the temporal scaling of correlation lengths and is in accord with theoretical expectations for a diffusive conserved order parameter~\cite{iengo2009renormalization}.

The most striking outcome of our work is the determination of the critical initial slip exponent $\theta = -1.052(5)$. This negative value stands in sharp contrast to the positive $\theta$ commonly found in conventional phase transitions described by the LGW paradigm. We attribute this anomalous behavior to two fundamental features of the dimer model at criticality: (i) the emergent SO(5) symmetry, which enhances fluctuation channels and suppresses early ordering, and (ii) the local U(1) gauge constraint (Gauss law), which imposes a conserved diffusive dynamics and slows down the early growth of the order parameter. The negative $\theta$ indicates that the order parameter undergoes an initial decay or remains suppressed before the onset of growth, a distinctive signature of the unconventional criticality present in this system.

Our study not only provides the first complete characterization of nonequilibrium short-time criticality in the 3D classical dimer model but also establishes a direct link between its exotic equilibrium universality class and distinctive dynamical scaling laws. The successful application of short-time dynamics in this context demonstrates its power as a robust numerical tool for probing critical phenomena in systems with complex constraints and emergent symmetries. In particular, these results may offer useful hints for understanding nonequilibrium dynamics near deconfined quantum critical points in related quantum systems, where emergent symmetries and gauge constraints also play an important role. For example, studying quantum matter using quantum computers has recently represented an emerging direction, wherein ground-state search relies on imaginary-time evolution. It was shown that imaginary-time evolution near the critical point obeys scaling similar to that of classical relaxation dynamics~\cite{PhysRevB.89.144115,PhysRevB.109.134309}. The work thus not only motivates us to further investigate the imaginary-time relaxation dynamics at the deconfined critical point, but also bears important implications for the experimental realization of novel quantum devices hosting similar deconfined critical points.



\begin{acknowledgments}
This project is supported by the National Natural Science Foundation of China (Grants No. 12222515 and No. 12075324), Research Center for Magnetoelectric Physics of Guangdong Province (Grant No. 2024B0303390001), the Guangdong Provincial Key Laboratory of Magnetoelectric Physics and Devices (Grant No. 2022B1212010008), and the Science and Technology Projects in Guangzhou City (Grant No. 2025A04J5408). ZY is supported by the Scientific Research Project (No.WU2025B011) and the Start-up Funding of Westlake University.
The authors also acknowledge the HPC Centres of SYSU and WLU and Beijing PARATERA Tech Co., Ltd., for providing HPC resources. 
\bigskip
\end{acknowledgments}



\bibliography{apssamp}

\end{document}